\documentclass[fleqn,usenatbib]{mnras}

\usepackage{newtxtext,newtxmath}
\usepackage[T1]{fontenc}
\usepackage{ae,aecompl}
\usepackage{graphicx}
\usepackage{amsmath}
\usepackage{amssymb}
\usepackage{comment}
\usepackage{array}

\newcommand\ergs{erg\,s$^{-1}$}
\newcommand\mdot{$\textrm{M}_\odot$}
\newcommand\mdotyr{$\textrm{M}_\odot$\,yr$^{-1}$}

\newcommand\kms{km\,s$^{-1}$}

\newcommand\cosmology{$H_0=70$ km\,s$^{-1}$\,Mpc$^{-1}$, $\Omega_M=0.3$, $\Omega_\Lambda=0.7$, $\Omega_k=0$}

\newcommand\magday{mag\,day$^{-1}$}
\newcommand\tigerfit{\href{https://github.com/manolis07gr/TigerFit}{TigerFit}}
\newcommand\cmcm{cm$^\textrm{2}$}
\newcommand\gcmcmcm{g\,cm$^{-3}$}
\newcommand\Ni{$^{56}$Ni}
\newcommand\Co{$^{56}$Co}

\title[SN 2008es at Late Times: NIR Excess and CSI]{The Type II Superluminous SN 2008es at Late Times: Near-Infrared Excess and Circumstellar Interaction}

\author[K. Bhirombhakdi et al.]{
Kornpob Bhirombhakdi,$^{1}$\thanks{E-mail: kb291313@ohio.edu}
Ryan Chornock,$^{1}$\thanks{E-mail: chornock@ohio.edu}
Adam A. Miller,$^{2,3}$ \newauthor
Alexei V. Filippenko,$^{4,5}$
S. Bradley Cenko,$^{6,7}$
and Nathan Smith$^{8}$
\\
$^{1}$Astrophysical Institute, Department of Physics and Astronomy, 251B Clippinger Lab, Ohio University, Athens, OH 45701, USA\\
$^{2}$Center for Interdisciplinary Exploration and Research in Astrophysics (CIERA),\\
and Department of Physics and Astronomy, Northwestern University, 2145 Sheridan Road, Evanston, IL 60208, USA\\
$^{3}$The Adler Planetarium, Chicago, IL 60605, USA \\
$^{4}$Department of Astronomy, University of California, Berkeley, CA 94720-3411, USA\\
$^{5}$Miller Senior Fellow, Miller Institute for Basic Research in Science, University of California, Berkeley, CA  94720, USA\\
$^{6}$Astrophysics Science Division, NASA Goddard Space Flight Center, Mail Code 661, Greenbelt, MD 20771, USA\\
$^{7}$Joint Space-Science Institute, University of Maryland, College Park, MD 20742, USA\\
$^{8}$Steward Observatory, University of Arizona, 933 N. Cherry Ave., Tucson, AZ 85721, USA \\
}

\date{Accepted XXX. Received YYY; in original form ZZZ}

\pubyear{2018}

\begin{document}
\label{firstpage}
\pagerange{\pageref{firstpage}--\pageref{lastpage}}
\maketitle

\begin{abstract}

SN 2008es is one of the rare cases of a Type II superluminous supernova (SLSN) showing no relatively narrow features in its early-time spectra, and therefore its powering mechanism is under debate between circumstellar interaction (CSI) and magnetar spin-down. Late-time data are required for better constraints. We present optical and near-infrared (NIR) photometry obtained from Gemini, Keck, and Palomar Observatories from 192 to 554 days after explosion. Only broad H$\alpha$ emission is detected in a Gemini spectrum at 288 days. The line profile exhibits red-wing attenuation relative to the early-time spectrum. In addition to the cooling SN photosphere, a NIR excess with blackbody temperature $\sim1500$ K and radius $\sim10^{16}$ cm is observed. This evidence supports dust condensation in the cool dense shell being responsible for the spectral evolution and NIR excess. We favour CSI, with $\sim2$--3 \mdot\ of circumstellar material (CSM) and $\sim$10--20 \mdot\ of ejecta, as the powering mechanism, which still dominates at our late-time epochs. Both models of uniform density and steady wind fit the data equally well, with an effective CSM radius $\sim 10^{15}$ cm, supporting the efficient conversion of shock energy to radiation by CSI. A low amount ($\lesssim 0.4$ \mdot) of \Ni\ is possible but cannot be verified yet, since the light curve is dominated by CSI. The magnetar spin-down powering mechanism cannot be ruled out, but is less favoured because it overpredicts the late-time fluxes and may be inconsistent with the presence of dust.

\end{abstract}


\section{Introduction} \label{sec:intro}

Superluminous supernovae (SLSNe), which generally look like normal supernovae (SNe) but are 10--100 times brighter at peak luminosity \citep[e.g.,][]{Gal-Yam2012}, have been discovered recently. Analogous to normal SNe (e.g., \citealt{alex1997}), SLSNe are classified as Type I for H-poor or Type II for H-rich. 
For Type II, SLSNe are subclassified into common cases of Type II with relatively narrow features (e.g., SN 2006gy; \citealt{Ofek2007.2006gy.observation,Smith2007.2006gy.observation}), and uncommon cases of Type II lacking such features (e.g., SN 2008es; \citealt{Miller2009,Gezari2009}).

The powering mechanisms of SLSNe are still under debate. Since SLSNe reach a peak ${\gtrsim} 10^{44}$ \ergs\ with total radiated energy ${\gtrsim} 10^{51}$ erg \citep{Gal-Yam2012}, if SLSNe are powered by \Ni, they requires $\sim1$--10 \mdot\ of \Ni, which is unrealistic for a core-collapse (CC) explosion.
Given a CC explosion supplying $\sim 10^{51}$ erg of shock energy, efficient conversion of the bulk kinetic energy into radiation by circumstellar interaction (CSI) is one natural explanation. However, some SLSNe, including Type I and Type II without narrow features, fail to show evidence supporting this explanation --- strong, relatively narrow (width of a few hundred to 1000 km s$^{-1}$) hydrogen emission lines analogous to those of typical SNe~IIn. Magnetar spin-down \citep{Kasen2010,Woosley2010} is currently the mainstream explanation for powering SLSNe which fail to support the efficient conversion. Other explanations include pair-instability (PI) explosion \citep{Gal-Yam2009pairInstability}, pulsational pair-instability (PPI) explosion \citep{Woosley2007pulsation,Woosley2017pulsation}, 
and fallback accretion \citep{Moriya2010fallback}.

SN 2008es is one of the rare cases of a SLSN~II without narrow features \citep{Miller2009,Gezari2009}. Other SLSNe~II lacking narrow features include SN 2013hx, OGLE-2014-SN-073, and PS15br \citep{Inserra2016,Terreran2017OGLE2014SN073}. The objects in this class have only very broad ($\sim 10,000$ \kms) H$\alpha$ emission, without a very narrow (unresolved) or relatively narrow (henceforth, ``narrow'') component. The powering mechanism of SN 2008es is still debated between CSI and magnetar spin-down. The early-time photometric data (up to $\sim$100 days) fit well to both models \citep{Chatzopoulos2013,Inserra2016}. Therefore, later-time data are required for better constraints.

Besides constraining the powering mechanism, the late-time data are also informative about the circumstellar environment, which in turn constrains the progenitor's evolution. Since the SNe cool down, at late times infrared emission from dust condensation is expected, and has been observed in many events, especially in typical SNe II \citep{Fox2011,Gall2011}. It is an interesting matter, which is still unknown, whether the dust emission can be similarly observed in SLSNe~II, as well as whether the dust component is newly condensed from the cooling SNe or is pre-existing in the circumstellar environment. SN 2006gy is one of the SLSNe~II which shows a near-infrared (NIR) excess as the sign of dust emission \citep{Smith2008a,Miller2010,Fox2015}. Here, we add SN 2008es onto the list.

In Section \ref{sec:data} of this paper, we present late-time photometric and spectroscopic data of SLSN 2008es from 192 to 554 days after explosion in the rest frame. The data include optical and NIR bands, which give the opportunity to investigate both powering mechanism and dust. In Section \ref{sec:ana}, we show the broad H$\alpha$ feature with red-wing attenuation, and also that there is a NIR excess, implying dust emission. Then, in Section \ref{sec:ana}, we try to explain the powering mechanism. We conclude in Section \ref{sec:conclusion}. Throughout, unless specified otherwise, all dates are UT, all SN phases are days after explosion in the rest frame, the assumed explosion date is MJD = 54574 and the peak of the light curve is MJD = 54602 \citep{Gezari2009}, all magnitudes are on the AB scale, the Galactic extinction is assumed to be $E(B-V) = 0.011$\,mag \citep{Schlafly2011.Galactic.extinction},
and the cosmology is \cosmology .

\section{Data} \label{sec:data}

\begin{table*}
	\centering
    \caption{Late-time photometry of SN 2008es}
    \label{table-photometry}
    \begin{tabular}{crcllccr}
		\hline
        Observation Date & \multicolumn{1}{c}{Phase} & Filter & Mag (observed) & Mag (corrected)$^a$ & SN Detection? & Telescope/ & \multicolumn{1}{c}{Exp. time}\\
        (UT) & \multicolumn{1}{c}{(days)} & & & & & Instrument & \multicolumn{1}{c}{(s)} \\
        \hline
        2008-12-05 & 192.12 & $i$ & 21.718 (0.068) & 21.800 (0.069) & Y & P200/COSMIC & 1530\\
        2009-02-18 & 254.36 & $K'$ & 23.494 (0.046) & 23.558 (0.049) & Y & Gemini/NIRI & 3120\\
        2009-02-19 & 255.19 & $V$ & (24.449) & (24.417) & N & Keck I/LRIS & 300\\
        2009-02-19 & 255.19 & $g$ & (25.776) & (25.737) & N & Keck I/LRIS & 420\\
        2009-02-19 & 255.19 & $R$ & 24.863 (0.298) & 25.270 (0.446) & Y & Keck I/LRIS & 390\\
        2009-02-19 & 255.19 & $I$ & 23.810 (0.192) & 23.928 (0.218) & Y & Keck I/LRIS & 300\\
        2009-04-16 & 301.66 & $H$ & 24.543 (0.189) & 24.768 (0.234) & Y & Gemini/NIRI & 3150\\
        2009-04-16 & 301.66 & $K'$ & 23.734 (0.118) & 23.816 (0.128) & Y & Gemini/NIRI & 1800\\
        2009-06-25 & 359.75 & $R$ & (25.092) & (25.066) & N & Keck I/LRIS & 1050\\
        2009-06-25 & 359.75 & $I$ & 24.439 (0.073) & 24.678 (0.095) & Y & Keck I/LRIS & 360\\
        2009-06-27 & 361.41 & $g$ & 26.436 (0.120) & (27.365) & N & Keck I/LRIS & 570\\
        2010-01-08 & 523.24 & $g$ & (25.570) & (25.531) & N & Keck I/LRIS & 1500\\
        2010-01-08 & 523.24 & $R$ & 25.123 (0.202) & 25.685 (0.352) & Y & Keck I/LRIS & 720\\
        2010-01-08 & 523.24 & $I$ & 24.765 (0.156) & 25.110 (0.220) & Y & Keck I/LRIS & 480\\
        2010-02-15 & 554.77 & $R$ & 25.698 (0.142) & 27.016 (0.527) & Y & Keck II/DEIMOS & 1020\\
        2010-02-15 & 554.77 & $I$ & (25.113) & (25.095) & N & Keck II/DEIMOS & 960\\
        2011-03-01 & 871.78 & $g$ & 26.565 (0.198) & (27.304) & N & Keck I/LRIS & 1930\\
        2011-03-01 & 871.78 & $R$ & (25.379) & (25.353) & N & Keck I/LRIS & 1180\\
        \hline        
	\end{tabular}
    \begin{flushleft}
    $^a$After extinction correction and host-galaxy subtraction.
    \end{flushleft}

\end{table*}

\begin{table}
	\centering
    \caption{Host emission of SN 2008es (no extinction correction)}
    \label{table-hostemission}
    \begin{tabular}{ccc}
    	\hline
    	Filter & Mag (measured)$^a$ & Mag (modelled)$^b$\\    
        \hline
    	$B$ & 26.96 (0.25) & 26.75 (0.08)\\
        $g$ & 26.44 (0.27) & 26.45 (0.08)\\
        $V$ & - & 26.05 (0.08)\\
        $R$ & 25.96 (0.20) & 26.07 (0.08)\\
        $I$ & - & 26.13 (0.08)\\
        $F160W/H$ & 26.85 (0.40) & 26.34 (0.08)\\
        $K'$ & - & 26.53 (0.08)\\
        \hline
    \end{tabular}
    \begin{flushleft}
    $^a$From \citet{Angus2016} and \citet{Schulze2016}.\\
    $^b$Uncertainties come only from the estimate of the normalisation constant.
    \end{flushleft}
\end{table}

\begin{figure*}
  \centering
  \hspace*{-0.45in}
  \includegraphics[width=0.54\textwidth,keepaspectratio, angle=90]{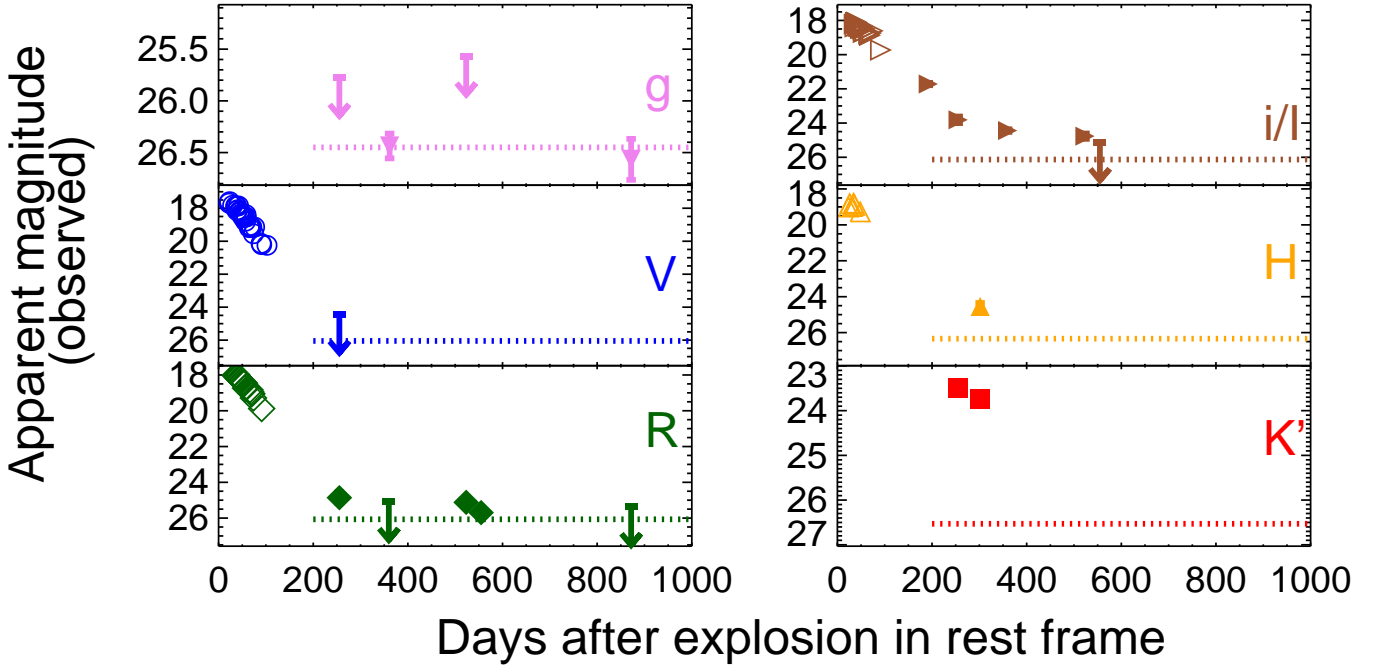}
  \caption{Photometry of SN 2008es in apparent magnitude. Filled symbols are the late-time data presented in this paper, while open symbols are the early-time data from \citet{Gezari2009} and \citet{Miller2009}. Diamond (purple) = $g$, square (blue) = $V$, circle (green) = $R$, upward triangle (brown) = $I$ (note that at 192 days the data is $i$ band before any correction), downward triangle (orange) = $H$, rightward triangle (red) = $K'$, downward arrow = upper limit, dotted horizontal line = modelled host-galaxy emission. The figure shows that the emission in $gVR$ converges to the host-galaxy light, while $IHK'$ is significantly brighter because of the strong H$\alpha$ emission in the $I$ band and the NIR excess in the $HK$' bands.}
  \label{fig-photo}
\end{figure*}

SN 2008es is located at $\alpha = 11^{h}56^{m}49.13^{s}$, $\delta = +54^\circ 27' 25.7''$ (J2000.0) at redshift $z=0.205$ \citep{Gezari2009}. Our late-time observations include one epoch of H$\alpha$ spectroscopy, as shown in \autoref{fig-spectra}, and several epochs of optical and NIR photometry, as shown in \autoref{table-photometry}. Our late-time photometry covers 2008 December 5 (192 days) to 2010 February 15 (554 days), including one epoch from the Palomar 200-inch Hale telescope (P200) with the Carnegie Observatories Spectroscopic Multislit and Imaging Camera (COSMIC) in the $i$ band\footnote{http://www.astro.caltech.edu/palomar/observer/200inchResources/cosmicspecs.html}, several epochs of $gVRI$ imaging obtained with the Low Resolution Imaging Spectrometer (LRIS) on the Keck I 10~m telescope \citep{Oke1995.LRIS,Rockosi2010.LRIS.upgrade} and with the DEep Imaging Multi-Object Spectrograph (DEIMOS) on Keck II \citep{Faber2003.DEIMOS}, and two epochs of $HK'$ from the Near InfraRed Imager and spectrograph (NIRI) on Gemini \citep{Hodapp2003.NIRI}. Additionally, we acquire $gR$ photometry from the public Keck Observatory Archive (KOA), extending the coverage to 2011 March 1 (871 days). 

We obtained a single 2000~s spectroscopic exposure using the Gemini Multi-Object Spectrograph \citep[GMOS;][]{Hook2004.GMOS} on the 8~m Gemini-North telescope on 2009 March 31.5 (288 days). Our instrumental setup used the R400 grating and a 1$\farcs$0 wide slit to cover the observed spectral range of 5500--9750~\AA\ at a resolution of 7~\AA. We used standard IRAF\footnote{IRAF is distributed by the National Optical Astronomy Observatory, which is operated by the Association of Universities for Research in Astronomy (AURA), Inc., under cooperative agreement with the National Science Foundation (NSF).} tasks to perform two-dimensional image processing and spectral extraction, as well as custom IDL routines to apply a relative flux calibration using an archival standard star. At the position of the transient, a very faint trace is barely detected in the continuum. However, a single broad emission feature is present between 7650--7950~\AA, which we identify as H$\alpha$ emission from the SN.

For photometric data, images of SN 2008es were reduced by following the standard procedures (bias, dark, flat, and photometric calibration) in IRAF. The data on 2011 March 1 were stacked from two different epochs to increase the signal-to-noise ratio (S/N): 2011 February 1 and 2011 March 26. Up to nine standard stars were identified in the field of images from the SDSS DR8 catalog for optical bands ($ugriz$), which were transformed to $UBVRI$ by following \cite{Blanton2007}. We calibrated the LRIS $g$-band images to SDSS $g$ (the two bands differ slightly). For NIR bands, the standard star FS 21 observed on 2009 April 16 was taken as the standard for calibrating $HK'$ at the same epoch, while $K'$ on 2009 February 18 was calibrated by creating a catalog from the stars in the field observed on 2009 April 16. The quality of the created catalog was verified with a few stars presented in the field of view and presented in the 2MASS catalog. For AB conversion, we followed \cite{Blanton2007}, \cite{Breeveld2011}, and \cite{Tokunaga2005}. For consistency with the other observations of SN~2008es, we transformed the $i$-band data from 2008 December 5 to the $I$ band using $I$(AB) = $i$(AB) $- 0.518$, found by assuming constant colour from 2009 February 19 with transformation equations from \cite{Blanton2007}. This was a reasonable assumption since SN 2008es converged to a temperature of $T = 5000$--6000~K by the end of the early-time observations \citep{Gezari2009,Miller2009}. 

\autoref{table-photometry} shows the observed AB magnitudes for the source at the position of the SN, including contamination from the host galaxy and the Galactic extinction. Some data are marked non-detection because their fluxes are less than $3\sigma$ above zero; these data are reported as $3\sigma$ upper limits (in parentheses). \autoref{fig-photo} plots the late-time data, including the earlier-time data from \cite{Gezari2009} and \cite{Miller2009}. 

Next, a faint ($M_{R} \approx 26$\,mag) host galaxy has been previously reported \citep{Angus2016}. The late-time data tend to converge to constants, corresponding to the host emission. Host subtraction was performed numerically owing to the lack of template images in several filters and the low significance of several of the detections, including those of the host only. A Galactic extinction correction was applied. Host-galaxy extinction was assumed to be negligible because the host of SN~2008es is blue and has low metallicity \citep{Angus2016,Schulze2016}, and then host subtraction was performed via adopting a host-galaxy model from \href{http://www.stsci.edu/science/starburst99/docs/default.htm}{Starburst99} \citep{Leitherer1999.starburst99,Leitherer2010.starburst99,Vazquez2005.starburst99,Leitherer2014.starburst99}. These templates, which we believe to be a good representative for the host of SN 2008es, are simulated for an instantaneous burst of star formation given the initial mass function with power-law index 2.35 over the range 1--100 \mdot, and nebular emission included. The templates include metallicity 0.001--0.04 and age 1--900\,Myr. The best galaxy model was selected by fitting the measured $BgR/F160W$ emission of the host of SN 2008es from \cite{Angus2016} and \cite{Schulze2016}, as shown in \autoref{table-hostemission}. We note that the host images in the bands $BR$ and $F160W$, which is equivalent to the $H$ band, were taken at phase $\sim 1700$ days, much later than the last $H$ data presented in \autoref{fig-photo}. We assume that there is no SN contamination at $\sim 1700$ days. The best-fit galaxy, determined by the lowest summed squared residuals, has metallicity 0.001 and an age of 200 million years, which is consistent with the results of \cite{Schulze2016}. Then, the host emission estimated from the best-fit galaxy model was estimated for each band, as shown in \autoref{table-hostemission}; \autoref{fig-photo} also shows the modelled host emission. We note that the estimated uncertainties of the modelled emission are unrealistically low. This is because only the statistical error from estimating the normalisation factor is included. However, as we will see, our analysis is insensitive to this.

Then, we apply the modelled host emission to perform the host subtraction. \autoref{table-photometry} shows corrected AB magnitudes of the late-time data after extinction correction and host subtraction. We also note that in this column the $i$ data are also transformed into the $I$ band. Some data, which are detections before the correction, are marked as nondetections because the corrected fluxes are less than $1\sigma$ above zero; therefore, these data are reported as $3\sigma$ upper limits (in parentheses). For some data which are marked as non-detection before the subtraction, only the extinction correction is applied, and the data are reported as $3\sigma$ upper limits. For a quick summary, \autoref{table-photometry} provides the column determining whether the data after the correction are considered as a SN detection.

\section{Analysis and Discussion} \label{sec:ana}

In this section, we analyse the data of SN 2008es and discuss the implications. First, we look at the H$\alpha$ emission, which exhibits a sign of dust condensation in the cool dense shell (CDS) and strong CSI but still shows no sign of narrow absorption/emission features. Then, we show that there exists a NIR excess corresponding to the thermal dust emission in the CDS. Last, we verify that CSI is the preferred powering mechanism, which is still the dominant mechanism during the late-time epochs.

\subsection{Spectroscopy: Strong CSI and CDS Dust Condensation}

\begin{figure}
  \centering
  \hspace*{-0.4in}
  \includegraphics[width=0.3\textwidth, angle=90]{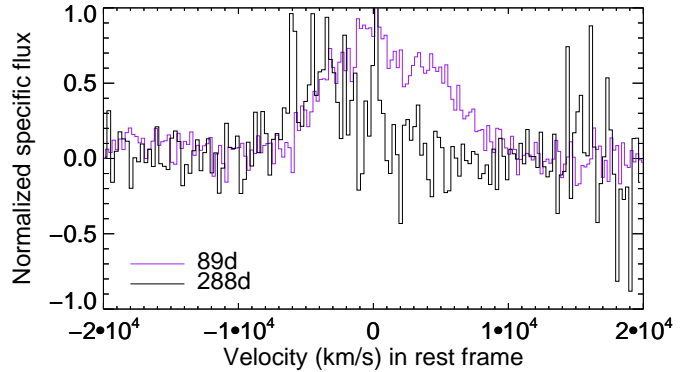}
  \caption{SN 2008es spectra, centered at H${\alpha}$, at 89 (purple) and 288 (black) days after explosion in the rest frame. A linear continuum has been subtracted from each spectrum to isolate the line emission.}
  \label{fig-spectra}
\end{figure}

It is common in SNe II that the strong CSI leads to the formation of a CDS, and dust condensation in this region at early times (i.e., ${\lesssim} 500$ days, which is earlier than the expectation of dust forming in the inner ejecta) \citep{Andrews2010,Andrews2016,Pozzo2004.1998S,Smith2008.2006jc,Smith2008sn2006tf,Smith2009,Smith2012.2010jl,Fox2011,Gall2011,Stritzinger2012,Gall2014.2010jl}. CDS is the region in between the forward and reverse shocks. A thermal instability can develop in this region, resulting in efficient cooling and becoming more dense \citep{Chevalier2017.csi}. Additionally, the cooling is also enhanced by metals in the ejecta. Because of this, dust condensation is very likely, and has been observed in many events \citep{Andrews2010,Andrews2016,Fox2011}.

To confirm dust condensation in the CDS, multiple pieces of evidence should be observed consistently. These include strong CSI, infrared excess, red-wing attenuation of spectral features, and the early onset (i.e., ${\lesssim} 500$ days) of these observational features \citep{Andrews2010,Andrews2016,Fox2011,Gall2011}. In this following sections \ref{sssec:strongcsi} and \ref{sssec:blueshifted}, we verify some of the mentioned evidence implied by the 288-day strong and blueshifted H$\alpha$ emission, as shown in \autoref{fig-spectra}.

\subsubsection{H$\alpha$ emission and strong CSI} \label{sssec:strongcsi}

From the photometry, \autoref{fig-photo}, we note the excess flux in the $I$ band. \autoref{fig-nirexcess} shows the excess relative to the assumed continuum of a 5000 K blackbody scaled to $R$ band. We assume the continuum blackbody temperature of 5000 K because there is evidence from the early-time analysis \citep{Gezari2009,Miller2009} that the temperature was converging to this value, which corresponds to the temperature of hydrogen recombination.

The excess $I$-band flux comes from strong line emission, as presented in \autoref{fig-spectra}. The figure shows spectra from the bandpass equivalent to the $I$ band. We clearly see the strong H$\alpha$ line emission.

The strong H$\alpha$ emission implies strong CSI. We can quantitatively show this by estimating the luminosity of H$\alpha$ emission and its equivalent width (EW). For the luminosity of H$\alpha$, since we cannot estimate directly from the spectra owing to the lack of an absolute calibration, we apply photometric data at 255 days instead. The $I$-band data have contributions from both the H$\alpha$ emission and the continuum, so we subtract the assumed continuum of a 5000-K blackbody scaled to the $R$ band, as presented in \autoref{fig-nirexcess}. The estimate yields $\sim 5 \times 10^{40}$ \ergs\ of H$\alpha$ emission at 255 days; at a similar epoch, this is comparable to some SNe IIn (e.g., SN 1988Z, \citealt{Turatto1993.1988Z}; SN 1998S, \citealt{Mauerhan2012.1998S.14yr})
and to SLSN-II with narrow features (e.g., SN 2006gy, \citealt{Smith2010}).

We can estimate the EW of the H$\alpha$ emission directly from the spectra. The estimate is 807 \AA\ at 288 days, and 161 \AA\ at 89 days. We note that, relative to the continuum estimated from the vicinity around the emission, H$\alpha$ emission at 288 days is significantly stronger than that of 89 days. At similar epoch, the 288-day EW is comparable to those of SN 1988Z (Type IIn) \citep{Stathakis1991.1988Z,Turatto1993.1988Z} and SLSN 2006tf (SLSN~II with narrow features) \citep{Smith2008sn2006tf}, and significantly stronger than that of SLSN 2006gy \citep{Smith2010}. The increasing trend of EWs of H$\alpha$ emission is also common in SNe~IIn, which are powered by CSI, even though SLSN 2006gy does not show such a trend \citep{Smith2008sn2006tf,Smith2009,Smith2010}.

\subsubsection{Blueshifted H$\alpha$ and CDS dust condensation} \label{sssec:blueshifted}

Red-wing attenuation of spectral features is expected, but not always, if dust is formed in the CDS \citep{Andrews2010,Andrews2016,Fox2011,Gall2011}. Observationally, the red-wing attenuated spectra show blueshifted peaks, and asymmetry by having the red-side emission weaker than that of the blue side, because dust in the CDS obscures more of the emission from the far side than from the near side. Progressively stronger attenuation with time is also expected because more dust is formed.

\autoref{fig-spectra} compares the shape of the 89-day and 288-day spectra of H$\alpha$ emission. The blueshifted peak in the 288-day spectrum is obvious, while the maximal velocity of the blue wing at $\sim 10,000$ \kms\ is similar to that of the 89-day one. This evidence, together with strong CSI and the early onset (i.e., as early as less than 288 days), supports the interpretation of dust condensation in the CDS.

Last, we note two other possible scenarios causing the observed blueshifted peak. First is the asymmetry of the ejecta with a higher concentration of the radioactive material (i.e., \Co\ during these epochs) toward the near side along the line of sight yielding more excitation and, therefore, more emission from the blue wing \citep{Hanuschik1988,Elmhamdi2003,Gall2011}. However, this is unlikely because \Co\ is not significantly powering the light curve (see Section \ref{ssec:poweringmechanism}). Second is asymmetry of CSM, with a higher concentration of CSM toward the near side of the ejecta enhancing the blue-wing emission. This scenario cannot be ruled out but is less favoured, compared to the interpretation of CDS dust, because the scenario does not explain the observed NIR excess. We show the evidence of a NIR excess and discuss its implication in the next section.

\subsection{NIR Excess: CDS Dust Emission} \label{ssec:nirexcess}

\begin{figure}
  \centering
  \hspace*{-0.1in}
  \includegraphics[width=0.475\textwidth, angle=90]{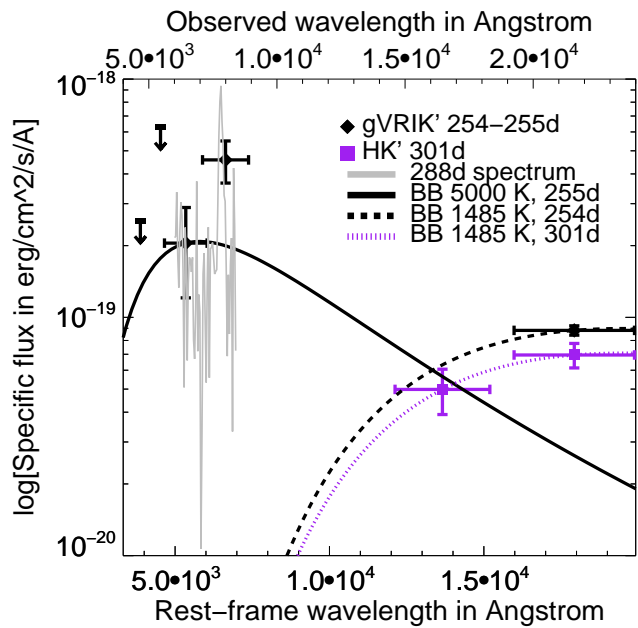}
  \caption{NIR excess. Data points are $gVRIK'$ (black, diamond) at 254--255 days, and $HK'$ (purple, square) at 301 days. Solid grey line = 288-day spectrum scaled to the $R$ band, showing H$\alpha$ contamination in the $I$ band. Solid black line = 5000~K blackbody, optical component, fit to the $R$ data at 255 days. Dashed black line = 1485~K blackbody, NIR component, scaled to the $K'$ data at 254 days. Dotted purple line = 1485~K blackbody, NIR component, fit to the $HK'$ data at 301 days. Downward black arrow = $3\sigma$ upper limit of the $gV$ bands at 255 days.}
  \label{fig-nirexcess}
\end{figure}

\begin{table}
	\centering
    \caption{Bolometric luminosity of the NIR component}
     \label{table-lumnir}
     \begin{tabular}{lclc}
     \hline
     \multicolumn{1}{c}{Phase} & \multicolumn{1}{c}{$\log_{10}[L]$} & \multicolumn{1}{c}{Temperature} & \multicolumn{1}{c}{Radius} \\
     \multicolumn{1}{c}{(days)} & \multicolumn{1}{c}{(erg\,s$^{-1}$)} & \multicolumn{1}{c}{(K)} & \multicolumn{1}{c}{(cm)} \\
     \hline
     254.36 & 41.59 (0.45) & 1485$^a$ & $1.06 \times 10^{16}$\\
     301.66 & 41.49 (0.45) & 1485 (218) & $9.41 \times 10^{15}$\\
     \hline
     \end{tabular}
     \begin{flushleft}
     $^a$Assumed 1485~K from 301 days.
     \end{flushleft}
\end{table}

In this section, we continue verifying the existence of the CDS dust by exploring the photometry. We discuss the evidence of a NIR excess, which is consistent with the interpretation of CDS dust condensation. \autoref{fig-photo} gives us a clue of the NIR excess by having emission in the $K'$ band brighter than if the emission came from the same continuum as the optical bands.

To be more specific, we present $gVRIK'$ data at 254--255 days and $HK'$ data at 301 days in \autoref{fig-nirexcess}. We note that $gV$ data are nondetections and the 301-day $HK'$ data are not contemporaneous with the optical $gVRIK'$ data -- about 50 days difference. To show the NIR excess, we fit the $gVRIK'$ data at 254--255 days with two blackbody components: optical and NIR. The optical $gVRI$ component is assumed to have $T = 5000$~K, implied by the temperature evolution shown in the early-time analysis being consistent with the temperature of hydrogen recombination \citep{Gezari2009,Miller2009}, and scaled to the $R$ band (because $gV$ are nondetections and $I$ is contaminated by H$\alpha$ emission). For the NIR $K'$ component, since we cannot fit the blackbody function nor the temperature, we approximate by fitting the component from the $HK'$ data at 301 days, and assume the same temperature in the range 254--301 days. We note that the contribution of the optical component at 301 days to the NIR component at the same epoch seems to be insignificant; we verify this by estimating the optical component at 301 days from assuming the same 5000~K blackbody temperature scaled to $R$ at 301 days estimated by the linear interpolation of the $R$ data between 255 and 523 days. The NIR component has a blackbody temperature of 1485~K. 

As shown in \autoref{fig-nirexcess}, the NIR excess component is obvious. The NIR excess about a year after the explosion supports the existence of thermal dust emission \citep{Fox2011,Gall2011}.

Next, we provide supporting evidence that the dust emitting this NIR excess is the CDS dust by showing that, first, the photospheric radius of the NIR component is located around the CDS region, and second, the radius is inconsistent with alternative explanations associated with CSM dust.

With the 1485~K temperature, we estimate the bolometric luminosity of the NIR component, shown in \autoref{table-lumnir}, by simply integrating the blackbody function. The implied photospheric radius is $\sim 10^{16}$ cm. The radius corresponds to the location of the forward shock, assuming an expansion velocity of 10,000 \kms\ as implied by the spectra. The correspondence of the location of the forward shock and the NIR component strongly supports the hypothesis that the CDS dust is responsible for emitting the observed thermal NIR excess; this is similar to the NIR-emitting CDS dust observed in some events such as SN 2005ip (Type IIn) \citep{Fox2009,Graham1983}. Moreover, the $\sim 1500$~K temperature of the NIR component is reasonable for the dust-condensation temperature.

The observed NIR excess is inconsistent with other explanations involving CSM dust emission (e.g., collision of ejecta, \citealt{Graham1986}; IR echo, \citealt{Dwek1983}) because the blackbody radius of $\sim 10^{16}$ cm is significantly smaller than the size of the dust-free cavity, at $\sim 10^{17}$ cm for typical parameters of SLSNe. The size of the dust-free cavity $R_{\rm evap}$ created by the SN peak flash is estimated by \citep{Dwek1983}
\begin{eqnarray}
	R_{\rm evap} = (23 \, \textrm{pc})\left(\frac{\bar{Q}_{\rm evap} \, (L_{\rm peak}/\textrm{L}_\odot)}{{(\lambda}_d / {\mu}\textrm{m}) \, T_{\rm evap}^5}\right)^{0.5},
	\label{eq:dustfree}
\end{eqnarray}
where $\bar{Q}_{\rm evap}$ is the mean grain emissivity, $L_{\rm peak}$ is the peak luminosity, $\lambda_d = 2 \pi a$ ($a$ is the radius of dust grain), and $T_{\rm evap}$ is the dust-evaporation temperature in kelvins. By assuming typical parameters for graphite grains of $\bar{Q}_{\rm evap} = 1$, $a = 0.1$ $\mu$m, and $T_{\rm evap} = 1900$~K, the peak flash of SN 2008es creates a dust-free cavity of size $\sim 10^{17}$ cm. We also note that this value tends to be a lower limit, since the size is sensitive to the evaporation temperature which is significantly lower for other dust species, such as 1200~K for silicate grains \citep{Dwek1983,Dwek1985,Fox2009,Fox2010}. 

Our analysis is sensitive to only the warm dust that emits at NIR wavelengths; colder dust, which lies farther away (for example, in the CSM) might exist and emits at longer wavelengths via mechanisms such as an IR echo, which is observed in SLSN 2006gy at epochs similar to those of our late-time observations \citep{Miller2010,Fox2015}. However, the emission from cold dust, if it exists, does not affect our interpretation of the warm dust.

\subsection{Powering Mechanisms} \label{ssec:poweringmechanism}

In this section, we discuss possible powering mechanisms of SN 2008es, specifically CSI and magnetar spin-down \citep{Gezari2009,Miller2009,Chatzopoulos2013,Inserra2016}. Both candidates fit well with the early-time data, and can be constrained better by our later-time data. We start by discussing the evolution of the light curve in general. Then, we show that CSI is more preferred, and yields implications consistent with other observed evidence. However, we also show that magnetar spin-down cannot be ruled out (but is less favoured).

\subsubsection{Evolution of the light curve of SN 2008es}

\begin{table}
	\centering
    \caption{Bolometric luminosity of late-time optical component}
    \label{table-lumopt}
    \begin{tabular}{clcc}
    	\hline
    	\multicolumn{1}{c}{Phase} & \multicolumn{1}{c}{$\log_{10}[L]$} & \multicolumn{1}{c}{Temperature$^a$} & \multicolumn{1}{c}{Radius} \\
        \multicolumn{1}{c}{(days)} & \multicolumn{1}{c}{(erg\,s$^{-1}$)} & \multicolumn{1}{c}{(K)} & \multicolumn{1}{c}{($10^{14}$ cm)} \\
    	\hline
    	192.12 & 42.28--42.63$^b$ & 5000 & 20.7--31.0 \\
        255.19 & 41.43 (0.18) & 5000 & 7.76 \\
        359.75 & $<$41.51 & 5000 & $<$8.52 \\
        \hline
    \end{tabular}
    \begin{flushleft}
    $^a$Assumed to be 5000 K. \\
    $^b$See text for the estimation of lower and upper limits.
    \end{flushleft}
\end{table}

\begin{figure}
  \centering
  \hspace*{-0.35in}
  \includegraphics[width=0.3\textwidth, angle=90]{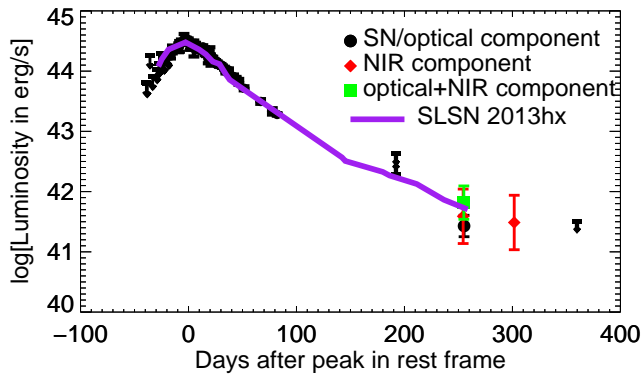}
  \caption{Bolometric luminosity of SN 2008es compared with SN 2013hx. Circle (black) = optical component, diamond (red) = NIR component, square (green) = optical + NIR component, downward arrow = $3\sigma$ upper limit, upward arrow = $3\sigma$ lower limit, solid line (purple) = bolometric luminosity of SN 2013hx \citep{Inserra2016}.}
  \label{fig-compare}
\end{figure}

The evolution of the light curve of SN 2008es is shown in \autoref{fig-photo} for each filter (which is discussed in the previous section), and in \autoref{fig-compare} for the bolometric luminosity including the early-time data from \cite{Gezari2009} and \cite{Miller2009} as well as our later-time data shown in \autoref{table-lumnir} and \autoref{table-lumopt}. When determining the bolometric luminosity, we estimate separately the NIR excess component from the SN component, so that we can investigate the contribution from each component. The bolometric luminosity of the SN component, which we refer to as the optical component, is estimated by simply integrating the 5000~K blackbody. At day 192, our only observation is in the $i$ band, which is potentially contaminated by H$\alpha$ emission. We set an upper limit by scaling the blackbody to the $I$ band, which is equivalent to assuming negligible H$\alpha$ contamination. The lower limit is estimated by assuming constant $R-I$ index from 255 days; this is set as a lower limit since the colour at 192 days can be bluer, hence brighter, than assumed if the EW of H$\alpha$ emission is increasing with time. At 359 days, the upper limit is estimated from the $3\sigma$ upper limit in the $R$ band. For the rising part, the data from ROTSE-IIIb in \cite{Gezari2009} are transformed into equivalent $R$-band points by using the data near peak; then, we assume constant temperature during the rise to the peak to estimate the bolometric luminosity. For the NIR excess component, we integrate the 1485~K blackbody function for the bolometric luminosity. 

The light curve has a peak $\sim 3 \times 10^{44}$ \ergs, and the estimated explosion is at about 23 days \citep{Gezari2009}. (Note that \autoref{fig-compare} shows days after peak brightness). Then, it linearly decays (in magnitude) until the end of the early-time data. At later times, the NIR component shows a slow decay rate of $0.005 \pm 0.003$ \magday, comparable to the rate of \Co\ decay at 0.01 \magday\ within $2\sigma$. This sets the upper limit of the initial $^{56}$Ni mass to $\lesssim 0.4$ \mdot\ by scaling the luminosity from $^{56}$Co decay to the NIR components. We note that the evolution of the optical component depends on whether the constraints from a single band ($i$) at 192 days are correct; if so, the decay rate slows down during 100--192 days, and then goes through another fast decay before slowing down again. 

In addition, \autoref{fig-compare} shows the bolometric light curve of SN 2013hx \citep{Inserra2016}, which is also a SN~II without narrow features. 
Although the light curves are strikingly similar, the spectral evolutions of the two objects differ, leading to different interpretations. While our spectra of SN 2008es show red-wing attenuation implying the existence of dust formation, the spectra of SN 2013hx exhibit H${\alpha}$ emission with multiple peaks and multiple velocity components, implying the interaction with asymmetric CSM \citep{Inserra2016}. This similarity in the light curves may imply similar powering mechanisms. At $\sim 300$ days after peak brightness, SN 2013hx shows brighter emission in the $K$ band relative to optical bands \citep{Inserra2016}, hinting the possible NIR excess. However, there is not enough information to verify this, and whether dust emission exists in SN 2013hx is an interesting question deserving of future investigation. Besides SN 2013hx, other SLSNe~II lacking narrow features include PS15br \citep{Inserra2016} and OGLE-2014-SN-073 \citep{Terreran2017OGLE2014SN073}. Their light curves differ from that of SN 2008es, thus implying possibly different powering mechanisms.

\subsubsection{CSI}

\begin{table*}
	\centering
    \caption{Fit results from CSMRAD model from \tigerfit}
    \label{table-csmrad}
    \begin{tabular}{crrrr}
    	\hline
    	Parameters & CSMRAD1 & CSMRAD2 & CSMRAD3 & CSMRAD4 \\
    	\hline
        data$^a$ & early & early + late & early & early + late \\        
        $s$ & 0 & 0 & 2 & 2 \\
    	$M_{\rm Ni}$ (\mdot) & 0.012 & 0.001 & 0.000 & 0.039 \\
        $E_{\rm SN}$ ($10^{51}$ erg) & 5.856 & 5.800 & 5.155 & 5.427 \\
        $R_{p}$ ($10^{14}$ cm) & 5.072 & 4.617 & 1.761 & 1.707 \\
        $M_{\rm ej}$ (\mdot) & 11.591 & 11.271 & 16.308 & 15.473 \\
        $\kappa_{\rm ej}$ (\cmcm g$^{-1}$) & 0.30 & 0.30 & 0.36 & 0.34 \\
        $d$ & 2 & 2 & 2 & 2 \\
        $n$ & 12 & 11 & 12 & 12 \\
        $M_{\rm CSM}$ (\mdot) & 2.668 & 2.349 & 2.647 & 2.491 \\        
        $R_{\rm CSM}$ ($10^{14}$ cm) & 12.759 & 11.672 & 15.574 & 13.417 \\
        $\rho_{\rm CSM}$ ($10^{-13}$ \gcmcmcm) & 6.544 & 7.519 & 98.249 & 116.138 \\
        Reduced $\chi ^2$ & 3.643 & 3.267 & 3.669 & 4.851 \\
        \hline
    \end{tabular}
    \begin{flushleft}
		$^a$Fit with early-time data, or including late-time data at 192 and 255 days. 
    \end{flushleft}
\end{table*}

\begin{figure}
  \centering
  \hspace*{-1.6in}
  \includegraphics[width=0.48\textwidth, angle=90]{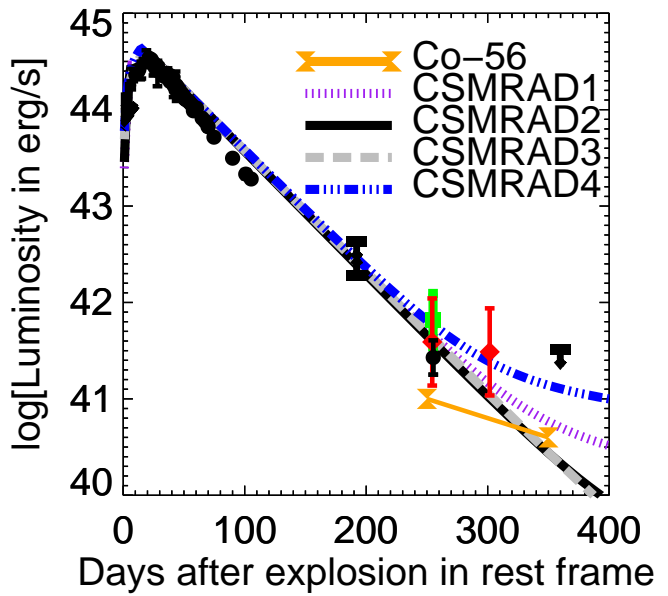}
  \caption{Bolometric luminosity of SN 2008es with models of CSI and \Ni\ powering. Circle (black) = optical component, diamond (red) = NIR component, square (green) = optical + NIR component, solid line with hourglass (orange) = \Co\ decay, dotted line (purple) = CSMRAD1, solid line (black) = CSMRAD2, dashed line (grey) = CSMRAD3, dot-dot-dot-dash line (blue) = CSMRAD4.}
  \label{fig-sborad}
\end{figure}

Efficient conversion of shock energy to radiation by CSI seems to be a natural explanation for the powering mechanism in SLSNe~II with narrow features, such as SLSN 2006gy \citep{Ofek2007.2006gy.observation,Smith2007.2006gy.observation}. Although SLSN 2008es lacks narrow features, its bolometric light curve fits well with the CSI model at early times. Here, we include our later-time data in a similar analysis for a better constraint on the mechanism.

We apply a semi-analytical model of CSI by using the CSMRAD routine in the \tigerfit\ package\footnote{https://github.com/manolis07gr/TigerFit}. 
Similar to \cite{Chatzopoulos2012}, \cite{Chatzopoulos2013}, and \cite{Wheeler2017tigerfitSN2017egm}, this model implements CSI with a diffusion process, including forward/reverse shock interaction, and radioactive (i.e., \Ni\ and \Co) heating. Parameters in the model include the initial \Ni\ mass $M_{\rm Ni}$, explosion energy $E_{\rm SN}$, progenitor radius $R_p$ (which is equivalent to the inner radius of the CSM in this model), ejecta mass $M_{\rm ej}$, ejecta opacity $\kappa_{\rm ej}$, power-law index of the density profile of the inner ejecta $d$ and of the outer ejecta $n$, power-law index of the density profile of the CSM $s$, CSM mass $M_{\rm CSM}$, mass-loss rate $\dot{M}$, and CSM wind velocity $v_w$. We note that, because of the large parameter set and nonlinearity of the model, the model tends to have high degeneracy which yields nonunique solutions with some uncertainty. Therefore, determining the best fit requires careful inspection.

\autoref{table-csmrad} shows four selected best-fit results. CSMRAD1 and CSMRAD3 are fed with only the early-time data, while the others also have the 192-day and 255-day (only optical component) data in the fit; we include the 192-day data by using the average and dispersion of the lower and upper limits. CSMRAD1 and CSMRAD2 assume a uniform density distribution ($s = 0$), while the others assume a steady wind ($s = 2$). To be comparable with the results of \cite{Chatzopoulos2013} and \cite{Inserra2016}, all models assume a power-law index of 2 ($d = 2$) for the density profile of the inner ejecta. We note that the solutions are insignificantly changed when applying $d = 0$, which is another common value used in the literature \citep{Wheeler2017tigerfitSN2017egm}. Also, we note that, in the table, we present the outer radius of the CSM $R_{\rm CSM}$ and the CSM density $\rho_{\rm CSM}$ instead of the mass-loss rate and the wind velocity by applying $\rho_{\rm CSM} = \dot{M}/(4 \pi v_w R_p^2)$ and $R_{\rm CSM} = [3M_{\rm CSM}/(4 \pi \rho_{\rm CSM} R_p^s) + R_p^{3-s}]^{1/3}$ (see \cite{Chatzopoulos2012} and also the code in \tigerfit). \autoref{fig-sborad} plots these models, showing that they are degenerate at early times but are distinguishable at later times. According to our coverage, we still cannot determine with certainty the best model among the four. It is interesting to note that, with only the early-time data, the solutions (CSMRAD1 and CSMRAD3) also fit well the later-time data, supporting the continuation of CSI dominating the light curve during the observational epochs.

The result of uniform-density models fitting with only the early-time data (CSMRAD1) is comparable to previous estimates \citep{Miller2009,Chatzopoulos2013,Inserra2016}. For all models, the results show similar properties of the progenitor and CSM. The estimate indicates a low mass of \Ni, implying that it is not the dominant source of energy during our observational epochs. The explosion energy is $\sim 5 \times 10^{51}$ erg with ejecta mass $\sim 10$--20~\mdot. The effective CSM mass is $\sim 2$--3~\mdot, which is comparable to that of SN 2006tf (superluminous SN~II with narrow features; \citealt{Smith2008sn2006tf}), but less than that of SN 2006gy having $\sim 10$~\mdot\ \citep{Miller2010}. Comparing to typical SNe~IIn, which have CSM mass $\sim 0.1$--10~\mdot\ \citep{Branch2017.supernovaExplosion.book}, the estimated CSM mass of SN 2008es is greater than that of SN 2005ip having $\sim 0.1$~\mdot\ \citep{Smith2009}, comparable to that of SN 2010jl \citep{Andrews2011}, but less than that of SN 1988Z having $\sim 10$~\mdot\ \citep{Aretxaga1999.1988Z}. The effective outer radius of the CSM is $\sim 10^{15}$~cm, comparable to the photospheric radius at peak brightness and supporting the efficient conversion mechanism. For the steady-wind models, the mass-loss rate is $\sim 0.1$--1~\mdotyr\ given a wind velocity of $\sim 100$ km\,s$^{-1}$, and for the uniform-density models the CSM density is $\sim 10^{-12}$--$10^{-13}$~\gcmcmcm. 

We investigate the potential radio emission properties of this CSI given the large derived mass loss rate of 0.1--1 \mdotyr\ and the explosion energy $\sim 5 \times 10^{51}$ erg estimated in the steady wind models following \cite{Chevalier1998.radio},\cite{Chevalier2006.radio}, and \cite{Soderberg2012.radio}, as synthesized by \cite{Coppejans2018.radio} and assuming similar microphysical parameters.  The synchrotron radio emission is heavily self absorbed at all early times when the shock is located within $R_{\mathrm{CSM}}$ derived above, but if the wind extends to a large radius we estimate the 5 GHz synchrotron radio emission to reach its peak at $\sim$1 mJy (i.e., ${\sim} 10^{30}$ erg/s/Hz) at an age of 6--20 years, corresponding to an interaction region at a radius of ${\sim} 10^{17}$ cm from the explosion site. However, it is unphysical for a steady wind with such a high mass loss rate to extend to this large radius without truncation because the total mass in the wind becomes very large, and so the true peak radio flux will lie below this estimate. Therefore, any prediction is uncertain because it depends on the CSM density at larger radii than those probed by the optical light curve presented in this work.

We note that the estimated mass-loss rate of SN 2008es is very high compared to known massive stellar winds, at most $\lesssim 10^{-3}$ \mdotyr\ with $v_w \approx 10$ km\,s$^{-1}$ for extreme red supergiants (RSGs) \citep{Smith2014.mass.loss,Vink2015.mass.loss,Smith2017.book.IIn.Ibn.mass.loss}. The mechanism for this extreme mass loss a few years before the explosion is still unknown, but is believed to be either by binary interaction ($\lesssim 10^{-1}$ \mdotyr\ with $v_w \approx 10$--100 km\,s$^{-1}$) or a luminous blue variable (LBV)-like giant eruption ($\lesssim 10$ \mdotyr\ with $v_w \approx 100$--1000 km\,s$^{-1}$) such as those observed in $\eta$ Carinae or P Cygni \citep{Smith2003.eta.car,Smith2006.PCygni.eruption,Smith2006.eruption.mass.loss,Smith2014.mass.loss,Smith2017.book.IIn.Ibn.mass.loss,Chevalier2012.mass.loss.binary}. The mass-loss rates of most of strong CSI events (such as SLSNe 2006gy and 2006tf, and SN~IIn 2010jl), are consistent with those of giant eruptions, while the mass-loss rates of some SNe~IIn (such as SNe 1988Z and 1998S) are consistent with those of binary interaction \citep{Smith2017.book.IIn.Ibn.mass.loss}. For SN 2008es, the estimated mass-loss rate is consistent with that of a giant eruption. Other proposed extreme mass-loss mechanisms include hydrodynamic instabilities \citep{Smith2014.mass.loss.instabilities}, gravity-wave-driven mass loss \citep{Shiode2014.mass.loss.gravity.wave}, or centrifugal-driven mass loss of spun-up Wolf-Rayet stars \citep{Aguilera-Dena2018.Wolf-Rayet}, which might be more related to the hydrogen-poor events rather than to the hydrogen-rich ones. 

Regardless of what exact mechanism causing the extreme mass loss, the CSM structure is unlikely to have a steady-wind profile, but is more likely approximated by a dense shell of uniform density \citep{Chatzopoulos2012}. Therefore, the CSI with wind models (i.e., CSMRAD3 and CSMRAD4) are less favourable, compared to the uniform-density ones. Also, the estimates assume spherical symmetry, yet it is likely that the CSM structure is actually complex. With bipolar/disc/torus shapes, multiple shells, or clumpy structure \citep{Andrews2010,Andrews2016,Smith2006.csm.sturcture.etaCar,Smith2009.clumpy.csm,Smith2014.mass.loss}, the mass-loss rate can be lower than that of spherical symmetry.

Our results strongly support CSI as the powering mechanism of SN 2008es. Moreover, the interpretation of CSI powering both the early-time and later-time emission is consistent with the high EW of H$\alpha$ and the existence of CDS dust, discussed in the previous section. 

Finally, we note that lacking narrow features, SN 2008es was argued to be inconsistent with the CSI powering scenario \citep{Gezari2009}. However, recent literature \citep{Smith2007,Woosley2007pulsation,Chevlier2011,Moriya2012,Smith2015.ptf11iqb,Andrews2018.iptf14hls} discusses how CSI powering SLSNe~II without narrow features is possible with some CSM configurations, such as a small amount of CSM mass or uniformly distributed material. Our results for SN 2008es are still uncertain regarding these suggested configurations, but as discussed previously, the CSM structure of SN 2008es is likely to be a uniform-density dense shell. Also, we can show by applying the analytical expressions of \cite{Chevlier2011} and \cite{Moriya2012} that SN 2008es matches the case without narrow features.

From \cite{Chevlier2011}, for a shock breaking outside the wind (i.e., CSI powering without narrow features), the duration of the rise is $R_w^2/(vR_d)$, and the duration before the rise is $R_w/v$, where $R_w$ is the radius of the CSM ejected by the steady wind and $R_d$ is the radius of the effective diffusion (i.e., breakout radius). Additionally, the duration from explosion to peak brightness is comparable to the effective light-curve timescale. Given the duration rise of 23 days, the expansion speed of 10,000 \kms, and the effective light-curve timescale of 66 days from the fit result of the toy model, we find that $R_w < R_d$ as required by this scenario. 

From \cite{Moriya2012}, we take the case of CSI powering without narrow features implied by $vt_{\rm LC}/R_0 \geq 1$, where $t_{\rm LC}$ is the effective light-curve timescale. Given $v = 10,000$ \kms\ and, for the most extreme case, $R_0 \approx 10^{15}$ cm (note that the largest known RSG has a radius of only $\sim 10^{14}$ cm), the condition is satisfied if $t_{\rm LC} \geq 11$ days (and this value is lower for a smaller progenitor radius). This is consistent with SLSN 2008es. Moreover, \cite{Moriya2012} mention that a uniform CSM distribution results in CSI powering without narrow features, and this is also consistent with the results from \tigerfit.

\subsubsection{Magnetar spin-down}

\begin{table}
	\centering
    \caption{Fit results from magnetar model$^a$}
    \label{table-mag}
    \begin{tabular}{lrr}
    	\hline
        Parameters & MAG1 & MAG2 \\
        \hline
        Trap$^b$ & O & I \\
        $t_{\rm LC}$ (days) & 19.47 (4.66) & 18.94 (2.40) \\
        $t_p$ (days) & 23.88 (19.96) & 23.92 (8.67) \\     
        $E_p$ ($10^{51}$ erg) & 2.41 (1.42) & 2.34 (0.58) \\
        $A$ (days$^2$) & 5424 (4576) & 5173 (1854) \\
        $P$ (ms) & 2.88 & 2.92 \\
        $B$ ($10^{14}$ G) & 1.28 & 1.30 \\
        $M_{\rm ej}$ (\mdot) & 0.53 & 0.50 \\
        $L(t=255)$ (erg\,s$^{-1}$) & $1.2 \times 10^{42}$ & $1.2 \times 10^{42}$ \\
        $L(t=302)$ (erg\,s$^{-1}$) & $7.1 \times 10^{41}$ & $6.9 \times 10^{41}$ \\
        $L(t=360)$ (erg\,s$^{-1}$) & $4.0 \times 10^{41}$ & $3.9 \times 10^{41}$ \\
        Reduced $\chi^2$ & 5.54 & 4.48 \\
        \hline
    \end{tabular}
    \begin{flushleft}
    $^a$Uncertainties in parentheses. \\
    $^b$Implementation of trapping function (O = outside integral, I = inside). \\
   
    \end{flushleft}
\end{table}

\begin{figure}
  \centering
  \hspace*{-0.35in}
  \includegraphics[width=0.3\textwidth, angle=90]{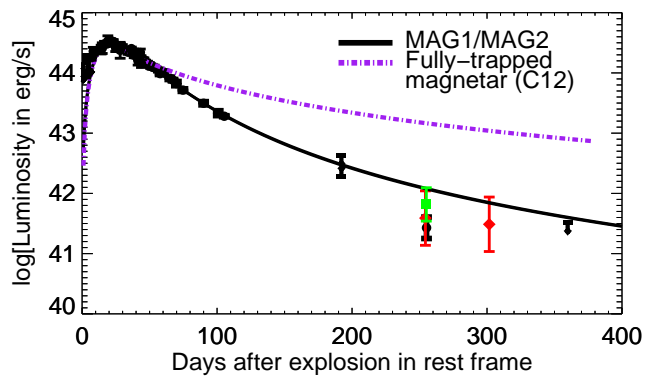}
  \caption{Bolometric luminosity of SN 2008es with magnetar spin-down model. Circle (black) = optical component, diamond (red) = NIR component, square (green) = optical + NIR component, solid line (black) = MAG1 and MAG2 (the lines overlap and cannot be distinguished), dot-dash line (purple) = fully-trapped magnetar spin-down fit from \citet{Chatzopoulos2012} implemented by \tigerfit.}
  \label{fig-mag}
\end{figure}

In this section, we fit the magnetar spin-down model to the light curve of SN 2008es. The model is \citep{Kasen2010,Woosley2010,Chatzopoulos2012}
\begin{eqnarray}
\label{eq-diffsmallradius} L(t) = \frac{2}{t_{\rm LC}} \exp \left[ - \frac{t^2}{t_{\rm LC}^2} \right] \times \\ \nonumber \int_{0}^{t} \exp \left[ \frac{t^2}{t_{\rm LC}^2} \right] \left( \frac{t}{t_{\rm LC}} \right) L_{\rm mag}(t) \, dt, \\
\label{eq-mag} L_{\rm mag}(t) = \frac{E_p}{t_p} \left( 1 + \frac{t}{t_p} \right)^{-2} \, ; \, t > t_p,
\end{eqnarray}
\noindent
where $L$ is the observed luminosity at time $t$ after the explosion powered by dipole-dominated magnetar spin-down with initial rotational energy $E_p$ and initial spin-down timescale $t_p$, passing through homologously expanding diffusive material with effective light-curve timescale $t_{\rm LC}$ and small initial radius. Additionally, we apply the trapping function $T = (1-\exp[-At^{-2}])$, where $A$ is the trapping coefficient and $A \rightarrow \infty$ for fully trapped energy \citep{Chatzopoulos2012,Chatzopoulos2013,Wang2015,Wang2016,Dai2016}. There are two different implementations for the trapping function, which we call case ``O'' for being outside the integral (i.e., $L = C T \int G dt$) and case ``I'' for being inside the integral (i.e., $L = C \int T G dt$), where $G$ is the integrand and $C$ is the multiplicative term in front of the integral. Physically, case ``O'' assumes that the bulk input luminosity is fully trapped during the diffusion process (i.e., $ C \int G dt$) but the observed luminosity is not, while case ``I'' assumes that the diffusion process cannot fully trap the input luminosity.

\autoref{table-mag} and \autoref{fig-mag} present the best-fit results by feeding only the early-time post-peak data; because of the condition $t > t_p$ as given in \autoref{eq-mag}, we omit the pre-peak data, and because of the uncertainty of the later-time conditions (e.g., changes in opacity and hard photon leakage) which are invalid for the model assumptions, we also omit the later-time data. Additionally, we plot the solution from \cite{Chatzopoulos2013} as the case of fully trapped energy for comparison purposes. This solution greatly overpredicts the late-time brightness.

The solutions MAG1/MAG2, which differ by the implementation of the trapping function but yield insignificantly different results, have the effective light-curve timescale $\sim 19$ days comparable to the spin-down timescale $\sim 24$ days, and have initial rotational energy $\sim 10^{51}$ erg. By applying Equations (1) and (2) of \cite{Kasen2010} and Equation (10) of \cite{Chatzopoulos2013}, the typical solution implies the magnetar with initial spin period $P \approx 3$ ms, field strength $B \approx 10^{14}$ G, and ejecta mass $M_{\rm ej} \approx 0.5$~\mdot.\footnote{We note that in the literature, there are slightly different definitions for calculating the spin-down timescale with different specifications. We follow the definition of \cite{Kasen2010}. See \cite{Nicholl2017slsnImagnetarMosfit} for a discussion of different specifications.} 
This solution is consistent with the SLSN magnetar described by \cite{Metzger2015}, and it is also consistent with results from other studies \citep{Kasen2010,Chatzopoulos2013,Inserra2016}. The solution fits the early-time data well, but predicts a brighter later-time light curve than what is observed; at 255 days, the discrepancy between the prediction of the typical solution and the observation is $\sim 5 \times 10^{41}$ \ergs, given that both the optical and NIR components are summed together. The discrepancy at late times is a common issue of fitting SLSNe with the magnetar model \citep{Kasen2010,Chatzopoulos2013,Wang2015,Inserra2016}. X-ray leakage or ionisation breakout is hypothesised to explain the discrepancy; however, besides SCP06F6 (SLSN I) showing very bright X-ray emission at early times \citep{Levan2013.SCP06F6.X-ray} and weak X-ray emission from SN~2006gy \citep{Smith2007.2006gy.observation}, there have been no other detections from the X-ray observations (especially in SLSNe I; \citealt{Margutti2017}).

Thus, we do not favour the magnetar model because the fit to the late-time observations is poor, and the magnetar scenario is likely incompatible with the observation of CDS dust, since the hot bubble produced by the magnetar is hostile to dust condensation \citep{Metzger2014}. However, we note that the magnetar scenario currently cannot be ruled out.

\section{Conclusion} \label{sec:conclusion}

We present and analyse late-time data (192--554 days after explosion in the rest frame) for SN 2008es, including optical/NIR photometry and spectroscopy of H$\alpha$. The spectra show strong and broad (without detected very narrow or relatively narrow components) H$\alpha$ emission with red-wing attenuation as early as 288 days, implying strong CSI and dust formation in the CDS. The blue-wing side of the emission extends to about 10,000 \kms, implying the ejecta expansion velocity being constant since the earlier-time data. The late-time photometry is consistent with a cooling SN photosphere and a NIR-excess component at $T \approx 1500$~K, implying thermal dust emission. The distance argument supports newly formed CDS dust being responsible for emitting the NIR excess, possibly heated by CSI. 

The analysis of the light curve supports CSI as the main powering mechanism from early times until the observed later-time epochs. The fit to the CSI model yields $\sim 10$--20 \mdot\ of ejecta and $\sim 2$--3 \mdot\ of CSM with either a uniform or steady-wind distribution. For the uniform-distribution model, the density is $\sim 10^{-13}$--$10^{-12}$ \gcmcmcm, while for the steady-wind model the mass-loss rate is $\sim 0.1$--1 \mdotyr\ given a wind velocity of $\sim 100$ \kms, consistent with that of an LBV-like great eruption. A uniform-density CSM shell is more likely than a stellar-wind structure. The effective CSM radius is $\sim 10^{15}$ cm, supporting the efficient conversion of shock energy to radiation by CSI. A low amount of of \Ni\ is estimated, $\lesssim 0.4$ \mdot\ (if excluding CSI) or 0.04 \mdot\ (if including CSI). The CSI powering scenario also provides a consistent explanation for the CDS dust condensation and strong H$\alpha$ emission. The magnetar spin-down powering mechanism cannot be ruled out, but it is less favourable because of the large brightness discrepancy at late times. Moreover, it is not consistent with other evidence at late times such as the NIR excess and strong CSI.

We note some limitations in our analysis. (1) The assumption of spherical symmetry of the CSM might not be valid, given the growing evidence supporting asymmetric or clumpy CSM \citep{Chugai2005,Fox2009,Fox2010,Fox2011,Kotak2009,Andrews2010,Andrews2011,Andrews2016,Tinyanont2016}. If this is the case, the interpretation of the condensation of the CDS dust will need to be reconsidered. However, this should not affect our other interpretations including the CDS dust condensation, which is still supported by the NIR excess and additional arguments. (2) The NIR observation is sensitive to warm dust, which corresponds to the CDS dust in our case. Colder dust located beyond the CDS could exist, and its emission might contaminate the NIR observation. If this is the case, we overestimate the NIR contribution to the energy budget. However, this should not affect our interpretation of CDS dust. (3) The assumption of a blackbody might be invalid, especially at late times when line emission dominates in the nebular phase. This limitation can affect significantly the estimation of luminosity and temperature. (4) The diffusion approximation in both CSI and magnetar spin-down assumes spherical symmetry, homologous expansion, a centrally-concentrated energy source, and constant opacity. Whether these assumptions hold for the analysis at late times is still unknown.

This work reveals, to some extent, the nature of SLSNe~II lacking narrow features, a very rare class of which SN 2008es was the first object. We note two important aspects of the class that need to be studied: the powering mechanism and dust production. The powering mechanism tends to be explainable by efficient CSI better than by magnetar spin-down. However, whether SN 2008es is a good representative of the class or is a unique case is still unknown. More objects of a similar nature are required. Besides SN 2008es, SLSNe~II without narrow features besides include (for example) SN 2013hx, OGLE-2014-SN-073, and PS15br \citep{Inserra2016,Terreran2017OGLE2014SN073}. Investigating the late-time behaviour of these objects might shed some light on the subject, although this will be challenging since they are distant. X-ray and radio observations are recommended probes for the CSI, as observed in some SNe~IIn (e.g., SN 1998S, \citealt{Pooley2002.98S.99em.xray}; SN 2010jl, \citealt{Chandra2015.2010jl.xray}) and in superluminous SN 2006gy \citep{Smith2007.2006gy.observation}. To explore dust production, NIR to mid-IR observations are recommended probes for future objects, and should be attempted with the {\it James Webb Space Telescope}.

\section*{Acknowledgements}

K.B. and R.C. acknowledge partial support from National Aeronautics and Space Administration (NASA) grant 80NSSC18K0665.
Support for A.V.F.'s supernova research has been provided by the NSF, the TABASGO Foundation, the Christopher R. Redlich Fund, and the Miller Institute for Basic Research in Science (U.C. Berkeley). A.V.F.'s work was conducted in part at the Aspen Center for Physics, which is supported by NSF grant PHY-1607611; he thanks the Center for its hospitality during the supermassive black holes workshop in June and July 2018.

This study based on observations obtained at the Gemini Observatory (Program ID GN-2009A-Q-48, PI Miller), which is operated by the Association of Universities for Research in Astronomy, Inc., under a cooperative agreement with the NSF on behalf of the Gemini partnership: the NSF (United States), the National Research Council (Canada), CONICYT (Chile), Ministerio de Ciencia, Tecnolog\'{i}a e Innovaci\'{o}n Productiva (Argentina), and Minist\'{e}rio da Ci\^{e}ncia, Tecnologia e Inova\c{c}\~{a}o (Brazil).

Some of the data presented herein were obtained at the W. M. Keck Observatory, which is operated as a scientific partnership among the California Institute of Technology, the University of California, and NASA. The Observatory was made possible by the generous financial support of the W. M. Keck Foundation. We thank J. M. Silverman for assistance with some of the Keck observations, and D. A. Perley for obtaining some of the Keck I/LRIS images. 
We thank M. Kasliwal for obtaining the P200/COSMIC images.
Also, this research has made use of the Keck Observatory Archive (KOA), which is operated by the W. M. Keck Observatory and the NASA Exoplanet Science Institute (NExScI), under contract with NASA. 
The authors wish to recognise and acknowledge the very significant cultural role and reverence that the summit of Maunakea has always had within the indigenous Hawaiian community.  We are most fortunate to have the opportunity to conduct observations from this mountain. 

\bibliographystyle{mnras.bst}
\bibliography{references} 

\appendix

\bsp	
\label{lastpage}
\end{document}